\documentclass[showpacs,preprintnumbers,amsmath,amssymb]{revtex4} 
\usepackage{graphicx}
\usepackage{epstopdf}
\usepackage{epsfig}
\usepackage{dcolumn}
\usepackage{bm}
\usepackage{amsmath}
\usepackage{psfig}
\begin{document}
\title{Extraction of Structure Functions from Quasielastic Electron 
Scattering $(e,e')$ from Medium and Heavy Nuclei}
\author{K.S. Kim} 
\email{kyungsik@color.skku.ac.kr}
\affiliation{ 
BK21 Physics Research Division and Institute of Basic Science,
Sungkyunkwan University, Suwon, 440-746, Korea} 
\author{L.E. Wright}
\email{wright@ohiou.edu} 
\affiliation{  
Institute of Nuclear and Particle Physics, Ohio University,
Athens, OH 45701}

\begin{abstract}
Using a relativistic mean-field single particle knock-out model  
for $(e,e')$ reactions on nuclei, we investigate approximate treatments 
of Coulomb distortion effects and the extraction of longitudinal and 
transverse structure functions.  We show that an effective momentum 
approximation (EMA) when coupled with a {\it  focusing} factor 
provides a good description of the transverse contributions to the 
$(e,e')$ cross sections for electron energies above 300 MeV  on 
$^{208}$Pb.   This approximation is not as good for the longitudinal 
contributions even for incident electron eneriges above 1 GeV and 
if one requires very precise extraction of longitudinal and transverse 
structure functions in the quasielastic region it is necessary to 
utilize distortion factors based on a nuclear model and a more accurate 
inclusion of Coulomb distortion effects.
\end{abstract}

\maketitle

\section{Introduction}

There continues to be considerable theoretical and experimental interest 
in extracting longitudinal and transverse structure functions as a function 
of energy loss for fixed three momentum transfer for a range of 
nuclei \cite{jlab}.  
For low to medium electron energies (200 MeV $< E <$ 600 MeV) the $(e,e')$ 
cross section is significantly affected by the static Coulomb field of 
the target nucleus for $Z$ $>$ 20 and it is necessary to use some method of 
removing the so-called ``Coulomb distortion effects'' in order to 
investigate the underlying physical processes in quasielastic scattering.  
For a precise extraction of the longitudinal response even at incident 
electron energies of 2 GeV, some correction for Coulomb distortion 
effects is needed.  
It would be very appealing if this removal of Coulomb effects could be 
effected by shifting the value of the experimental energy or scattering 
angle so as to compensate for the Coulomb distortion \cite{kim5}.  
While it is quite clear that this cannot be done exactly, the question 
is can one find some approximate treatment of Coulomb distortion effects  that improves with increasing electron energy.  

In this paper, we investigate this question and report very good success 
in handling the Coulomb distortion effects in the transverse parts of 
the quasielastic cross section, but with less success in the longitudinal 
portion of the cross section unless we utilize a nuclear model of the 
transition current arising from the quasielastic knocking out of nucleons.

In order to frame the discussion, it is useful to note that in the plane 
wave Born approximation (PWBA), the inclusive $(e,e')$ cross section  
for electrons or positrons (assuming nuclear current conservation) is given by 
\begin{equation}
\frac{d^2\sigma}{d\Omega_e d\omega}= \sigma_{M} \left\{
\frac{Q^4}{q^4}  S_L(q,\omega) + \left[ \tan^2 \frac{\theta_e}{2} +
\frac{Q^2}{2q^2} \right]  S_T(q,\omega) \right\}, \label{pwsep}
\end{equation}
where $q_\mu ^2 = \omega^2-{\bf q}^2=-Q^2$ is the four-momentum
transfer, $\sigma _{M}$ is the Mott cross section given by
$\sigma_{M} = (\frac{\alpha }{2E_i} )^2  \frac{\cos^2
\frac{\theta_e}{2}}{\sin^4 \frac{\theta_e}{2}}$, and $S_L$ and $S_T$
are the longitudinal and transverse structure functions which
depend only on the momentum transfer $q$ and the energy transfer
$\omega$.   
As is well known, by keeping the momentum and energy
transfers fixed while varying the electron energy $E_i$ and
scattering angle $\theta_e$, it is possible to extract the two
structure functions with two measurements.   
The longitudinal and transverse structure functions in Eq.~(\ref{pwsep}) 
are squares of the Fourier transform of the components of the nuclear
transition current density integrated over outgoing nucleon angles.  
Explicitly, the structure functions for knocking out nucleons from 
a nucleus are given by
\begin{eqnarray}
S_{L}(q,\omega)&=&\sum_{\alpha_b s_p} {S_{\alpha_b}{\rho}_{p}}
 \int {\mid}N_{0}{\mid}^{2}d{\Omega}_{p} \\
S_{T}(q,\omega)&=&\sum_{\alpha_b s_p} {S_{\alpha_b}{\rho}_{p}}
 \int
({\mid}N_{x}{\mid}^{2}+{\mid}N_{y}{\mid}^{2})d{\Omega}_{p} \label{strut}
\end{eqnarray}
where the nucleon density of states ${\rho}_{p}={\frac {pE_{p}}
{(2\pi)^{2}}}$, $s_p$ is the spin-projection of the continuum nucleon, 
the $z$-axis is taken to be along ${\bf q}$, and ${S_{\alpha_b}}$ 
contains the spectroscopic and averaging factors for each bound orbital.  
The Fourier transform of the nuclear current $J^{\mu}({\bf r})$ is simply,
\begin{equation}
N^{\mu}=\int J^{\mu}({\bf r})e^{i{\bf q}{\cdot}{\bf r}}d^{3}r.
\label{fourier} \end{equation}
and the continuity equation has been used to eliminate the
$z$-component ($N_{z}$) via the equation $N_{z}=-{\frac {\omega} {q}}N_{0}$.  
As noted above, the PWBA calculation breaks down for
cases when the static Coulomb potential at the surface of the nucleus is 
not negligible when compared to the incident and outgoing electron energy.  
To set the scale, note that the Coulomb potential at the surface of 
$^{208}$Pb is about 20 MeV. 

Based on our previous investigations of inclusive quasielastic scattering 
and a new appreciation of a {\it  focusing} 
factor \cite{jin1,jin2,kim1,kim3,kim4,aste}, we investigate an improved 
effective momentum approximation with focusing (EMA-f) in this paper.  
Our goal is to seek a procedure for extracting the longitudinal and 
transverse structure functions from experimental data with minimal use 
of nuclear models.

\section{Coulomb Distortion Effects in Inclusive Quasielastic Scattering}
As discussed in a previous paper \cite{kim4}, we found a treatment of Coulomb 
distortion, labelled  
{\it approximate DW},  for $(e,e')$ from medium and heavy nuclei 
that agrees to within a few percent to a full DWBA partial wave analysis 
which includes the static Coulomb field of the target exactly in the 
electron (or positron) wavefunctions.  
In this approximation we define an $r$-dependent momentum for the incoming 
and outgoing electron wavefunctions, ${\bf p}'_{i,f}(r)= 
(p_{i,f}-\frac{1}{r}\int_0^rV_c(r) dr) {\bf\hat{p}}_{i,f}$, 
where $V_c(r)$ is the potential energy of the lepton moving in the 
static Coulomb field of the target nucleus.  
This implies an $r$-dependent momentum transfer  
${\bf q}^\prime(r)={\bf p}^\prime_i(r)-{\bf p}^\prime_f(r)$.  
The choice of an $r$-dependent momentum allows the distorted partial 
wave solutions of the Dirac equation with the static Coulomb potential 
present to be approximated by spherical Bessel functions with argument 
$x=p'(r) r$ quite well from the origin out to more than 3 times the 
nuclear radius $R$.  
An additional requirement of approximating Coulomb distorted waves is 
to incorporate the Coulomb scattering phase shifts into the problem.  
As discussed in previous papers \cite{ kim1, kim4} we were able to achieve 
this by fitting the phase shifts as a function of the square of the Dirac 
quantum number $\kappa$ for the incoming and outgoing electron energies 
for the nucleus under investigation and then to replace $\kappa^2$ by 
the classical angular momentum squared $({\bf r} \times {\bf p})^2$ 
so that  ``plane-wave-like'' lepton wavefunctions can be constructed 
that contains the effects of Coulomb distortion in the local momentum 
and the parametrized phase shifts which lead to what other authors refer 
to as {\it focusing}.
Using this ``plane-wave-like'' wavefunction, we obtained an approximate 
``M{\o}ller-like'' potential given by,
\begin{equation}
A^{appro. DW}_{\mu} ({\bf r})={\frac {4{\pi}e} {q^2-\omega^2}
}e^{i[{\delta}_{i}+{\delta}_{f}]}
e^{i(\Delta_i-\Delta_f)}  e^{i{\bf q}'(r){\cdot}{\bf r}}
{\bar u}_{f}{\gamma}_{\mu}u_{i} \label{apppot}
\end{equation}
where the phase shift parametrization is given by
\begin{equation}
{\delta}({\kappa^2})=\left[a_0 + a_2\frac{{\kappa}^{2}}{(pR)^{2}}\right]
e^{-{\frac{1.4{\kappa}^{2}}{(pR)^{2}}}}-{\frac{{\alpha}Z}{2}} 
\left( 1-e^{-{\frac{{\kappa}^{2}}{(pR)^{2}}}} \right)
{\ln}(1+{\kappa}^{2})  \label{newph}
\end{equation}
where $\kappa^2=({\bf r}{\times}{\bf p})^2$ and  
$p$ is the electron momentum (for the initial and final kinematics) 
and we take the nuclear radius to be given by $R=1.12A^{1/3}-0.86A^{-1/3}$.  
The two constants $a_0$ and $a_2$ are fitted to two of the elastic 
scattering phase shifts ($\kappa=1$ and $\kappa=Int(pR)+5$) for the 
incident and final electron energy.   
The parameter $\Delta=a[{\bf {\hat p}}{\cdot}{\hat r}]
{\bf L}^{2}$ denotes a small higher order correction to the electron wave 
number which we have written in terms of the parameter 
$a=-{\alpha}Z(\frac{16 MeV/c}{p})^{2}$.   
The potential in Eq. (\ref{apppot}) cannot be easily decomposed into 
a multipole expansion due to the angular dependence of the vector {\bf r} 
in the expression for the phase shifts.  
However, when combined with the nuclear transition current  density 
$J^{\mu}({\bf r})$ as in Eq. (\ref{fourier}),  the modified Fourier 
transform can be obtained by three dimensional integration since the 
volume is limited to a sphere with radius of 3-4 times the nuclear radius $R$.  
We confirmed that use of the potential given in Eq. (\ref{apppot}) 
reproduces the full DWBA results for the cross section very well for 
electron energies above 300 MeV and for momentum transfer greater than 
about 250 MeV.  
With this {\it approximate DW} four potential $A_\mu$ it is
straightforward to calculate the  exclusive $(e,e'p)$ cross sections and
modified structure functions.  
We showed \cite{kim3} that using this new phase shift we can
reproduce the full DWBA cross sections for $(e,e'p)$ from medium
and heavy nuclei very well.

Again as noted in our previous paper \cite{kim4}, this {\it approximate DW} 
potential which includes the local value of the  potential (i.e, a function 
of $r$) and the focusing effect due to the phase shifts is very time consuming 
for the inclusive $(e,e')$ reaction since we need to integrate over the solid 
angle of the outgoing nucleons.  
Thus, we proposed making further approximations for the inclusive process.  
In order to allow a straight-forward multipole expansion, we must remove 
any angular dependence from the phase shifts in Eq. (\ref{apppot}).   
We investigated various methods of achieving this and found that neglecting 
the phase shifts entirely, but including a {\it focusing factor} was 
sufficient for the transverse contribution.  
However, this was not sufficient for the longitudinal term so we chose 
to include some of the effects of the phase shifts by averaging over 
the angles of the vector {\bf r} in  the phase shift parametrization.  
With these further approximations we were able to write the $(e,e')$ cross 
section\cite{kim4} as
\begin{equation}
\left(\frac{d^2\sigma}{d\Omega_e d\omega}\right)^{ad-hoc}= \sigma_{M} \left \{
\frac{Q^4}{q^4}  S'_L(q',\omega) + \left[ \tan^2 \frac{\theta_e}{2} +
\frac{Q'^2}{2q'^2} \right]  S'_T(q',\omega) \right \} \label{ad-hoc}
\end{equation}
where the Fourier transforms of the transition current in $ S'_L$ and 
$S'_T$ are replaced by
\begin{eqnarray}
N^{ad-hoc}_{0}&=& \int \left({\frac {q'_{\mu}(r)} {Q}} \right)^{2}
\left({\frac {q} {q'(r)}} \right)^{2}
e^{i<\delta_i+\delta_f>}e^{i{\bf
q}'(r){\cdot}{\bf r}}J_{0} ({\bf r})d^{3}r \label{ad-hoc0}  \nonumber \\
{\bf N}^{ad-hoc}_{T}&=&\left({\frac {p_{i}'(0)} {p_{i}}} \right) \int e^{i{\bf
q}'(r){\cdot}{\bf r}}{\bf J}_{T}({\bf r}) d^{3}r. \label{ad-hocT}
\end{eqnarray}
where $<\delta_{i,f}>$ denotes an average of $\delta(\kappa^2)$ over the
angles of the vector ${\bf r}$.  
That is, the argument of the parametrization for the phase shifts in 
Eq. (\ref{newph}) is given by  $<\kappa_{i,f}^2>= <({\bf r} \times 
{\bf p}_{i,f})^2>=r^2 p^{2}_{i,f}(3-cos^2 \theta_{p_{i,f}})/4$.    
In addition, we fix the direction of ${\bf q}^\prime$ to be equal to 
the asymptotic momentum transfer, but we use 
$q^{\prime 2}(r)= p^{\prime 2}_i(r) 
+ p^{\prime 2}_f(r)-2p^\prime_i(r) p^\prime_f (r)cos {\theta_e}$ 
for the magnitude.   We confirmed that for
the kinematics under consideration this change is negligible.
Using a toy model and the full three dimensional integration for the 
longitudinal terms and by comparing to the full DWBA for the cross section 
we found that this so-called {\it ad-hoc} model works very well for 
$(e,e')$ on medium and heavy nuclei.  

However, use of an $r$-dependent momentum transfer requires a nuclear model 
to extract longitudinal and transverse structure functions.  
The challenge is to find some approximation with a constant shift of 
momentum transfer which approximates Coulomb distortion.  
One such approximation, referred to as the effective momentum approximation 
(EMA) replaces the $r$-dependent momentum for the incoming and outgoing 
lepton wavefunctions with a fixed value given by $p^\prime=p-V_c(R_c)$ 
where $R_c$ is usually taken to be equal to 0 or the nuclear radius $R$.    
Then one calculates the effective momentum transfer in terms of these momenta 
and the electron scattering angle.  
It is convenient to calculate $V_c(R_c)$ for a uniform charge distribution 
of radius $R=1.12A^{2/3}-0.86 A^{-1/3}$ containing a charge of $Ze$.  
The result for $R_c<R$ is simply, $V_c(R_c)=-\frac{Z\alpha}{2R}(3-R_c^2/R^2)$ 
for electrons.  
However, this approximation does not include the effect of the phase shifts 
which result in what other authors refer to as {\it focusing}.
As discussed in some detail by the Basel group \cite{aste}, the effects 
of focusing can be included by multiplying the potential arising from 
the EMA wavefunction  by the factor $p^\prime_i(0)/p_i$.  
We will refer to this approximation as EMA-f and we note that the $(e,e')$ 
cross section for EMA-f is simply this factor squared times the EMA cross 
section.  Thus, the EMA-f cross section is given by
\begin{eqnarray}
\left({\frac {d^2\sigma} {d\Omega_e d\omega}}\right)^{EMA-f}= 
\left({\frac {p'_i(0)} {p_i}}\right)^2 {\sigma_M} \left\{ 
{\frac {{Q'}^4} {{q'}^4}}  S_L (q',\omega) + 
\left[ \tan^2 {\frac {\theta_e} {2}} + {\frac {{Q'}^2} {2{q'}^2}} \right]  
S_T (q',\omega) \right\}. 
\label{ema-f}
\end{eqnarray}
The structure functions depend on the standard Fourier transforms of the 
nuclear transition current except that they are calculated as a function 
of $q^\prime$ rather than $q$.  
Note that  ${Q'}^2={q'}^2-\omega^2$.  Since the sign of $V_c$ changes for 
positron induced reactions, clearly one can shift $E_i$ for positron 
induced reactions as compared to electron induced reactions such that
$q'_{e_-}=q'_{e_+}$ and therefore the quantity 
$(\frac{d^2\sigma}{d\Omega_e d\omega})^{EMA-f} /(\sigma_M 
(\frac{p'_i(0)}{p_i})^2)$ is the same for electrons and positrons 
in this approximation.  
That is, the focusing factor must be removed before the overall structure 
functions are equal for electrons and positrons.  

\begin{figure*} 
\includegraphics{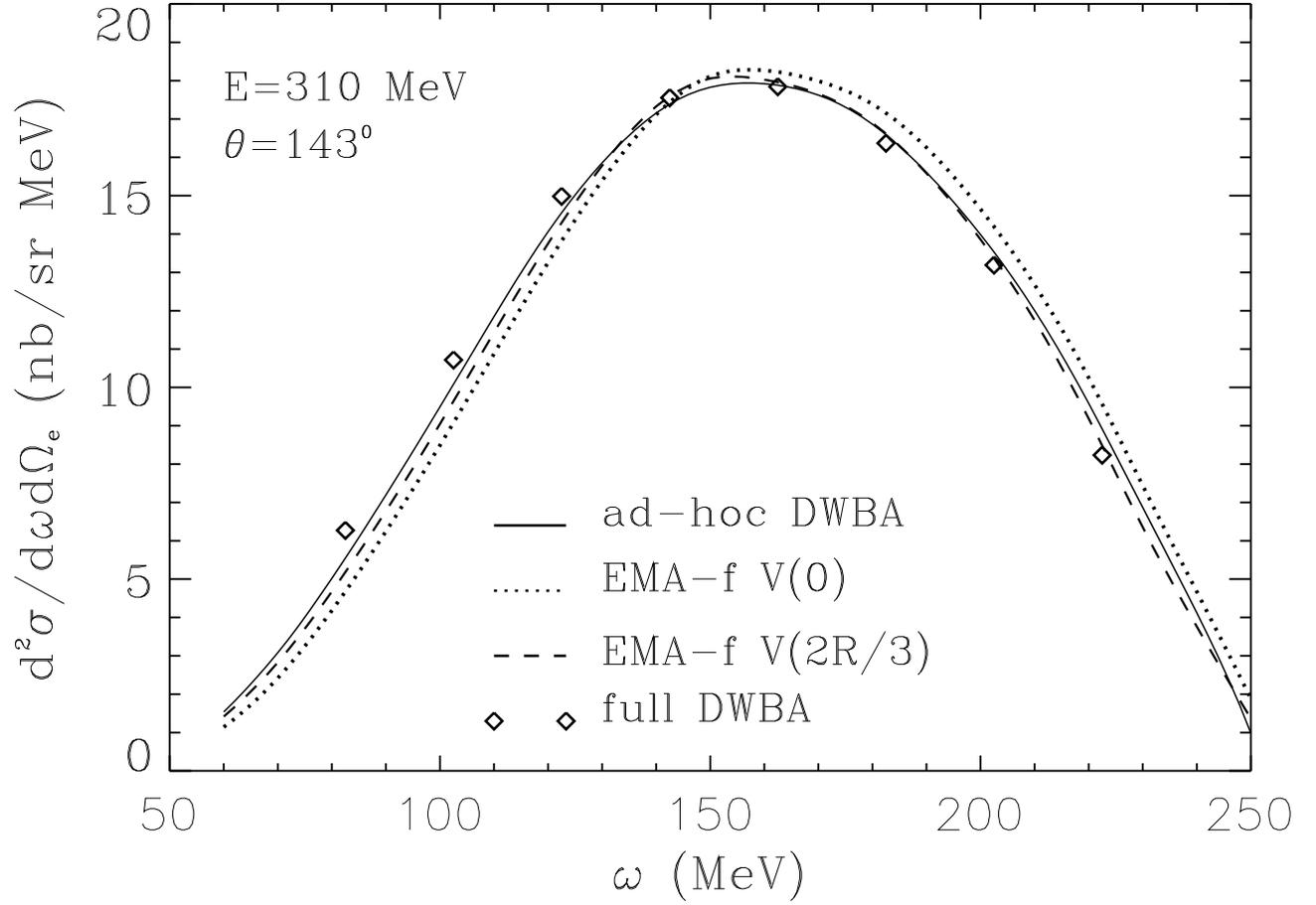} 
\caption{Theoretical quasielastic
scattering cross sections for $^{208}$Pb with incident electron
energy of $E_i$=310 MeV and electron scattering angle
$\theta=143^o$ as a function of energy transfer.  
The diamonds are the result of our full DWBA partial wave calculation.  
See text for details of the models and
for details of the {\it ad-hoc} and EMA-f results.}
\label{ema310}
\end{figure*}

\begin{figure*}
\includegraphics{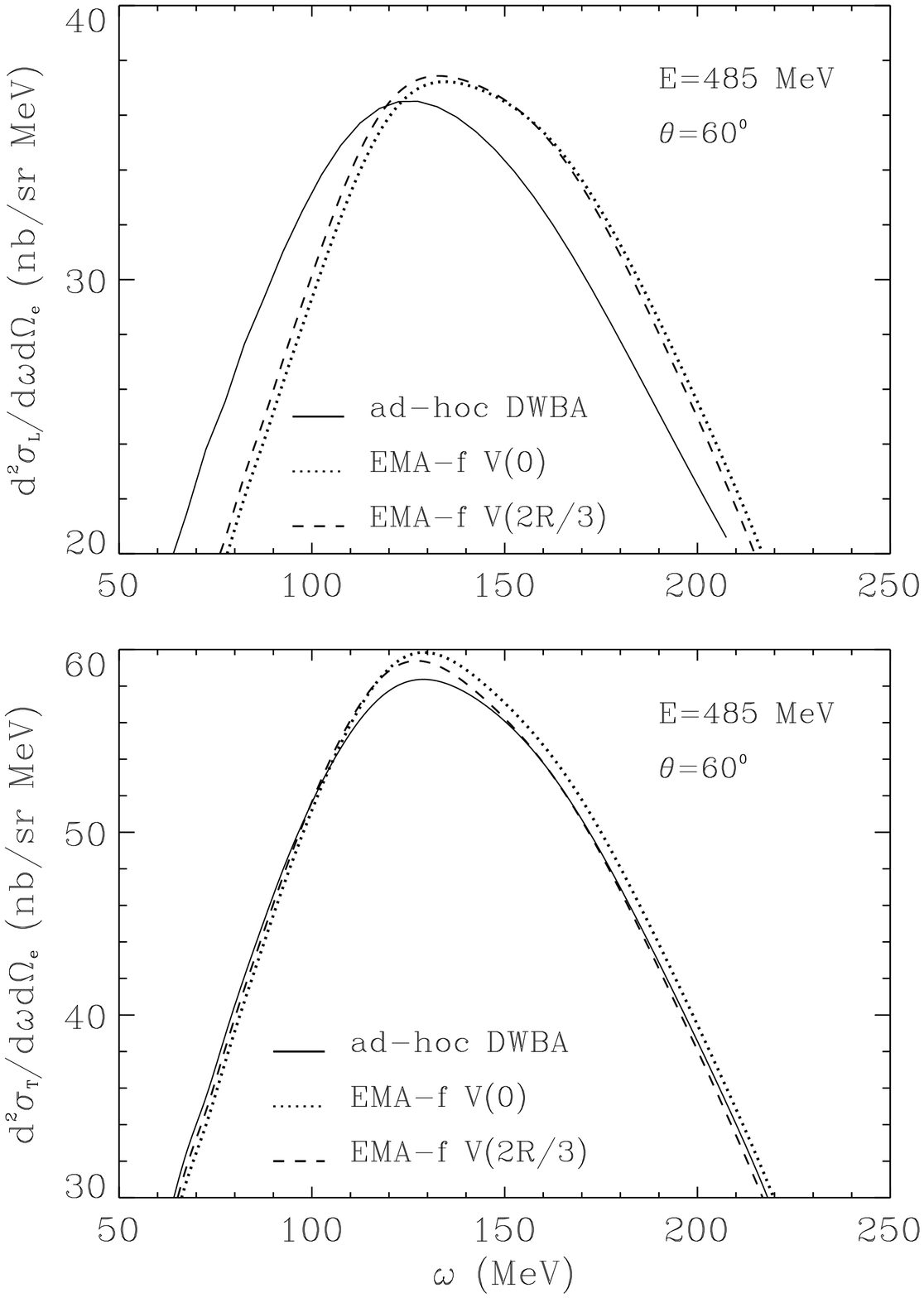} 
\caption{Theoretical quasielastic scattering partial cross sections 
for $^{208}$Pb with incident electron energy of $E_i$=485 MeV and electron 
scattering angle $\theta=60^o$ as a function of energy transfer.  
The upper panel shows the longitudinal contribution to the cross section 
while the lower panel shows the transverse contribution.  See the text 
for details of the {\it ad-hoc} and EMA-f calculations.}
\label{ss485}
\end{figure*}

\begin{figure*} 
\includegraphics{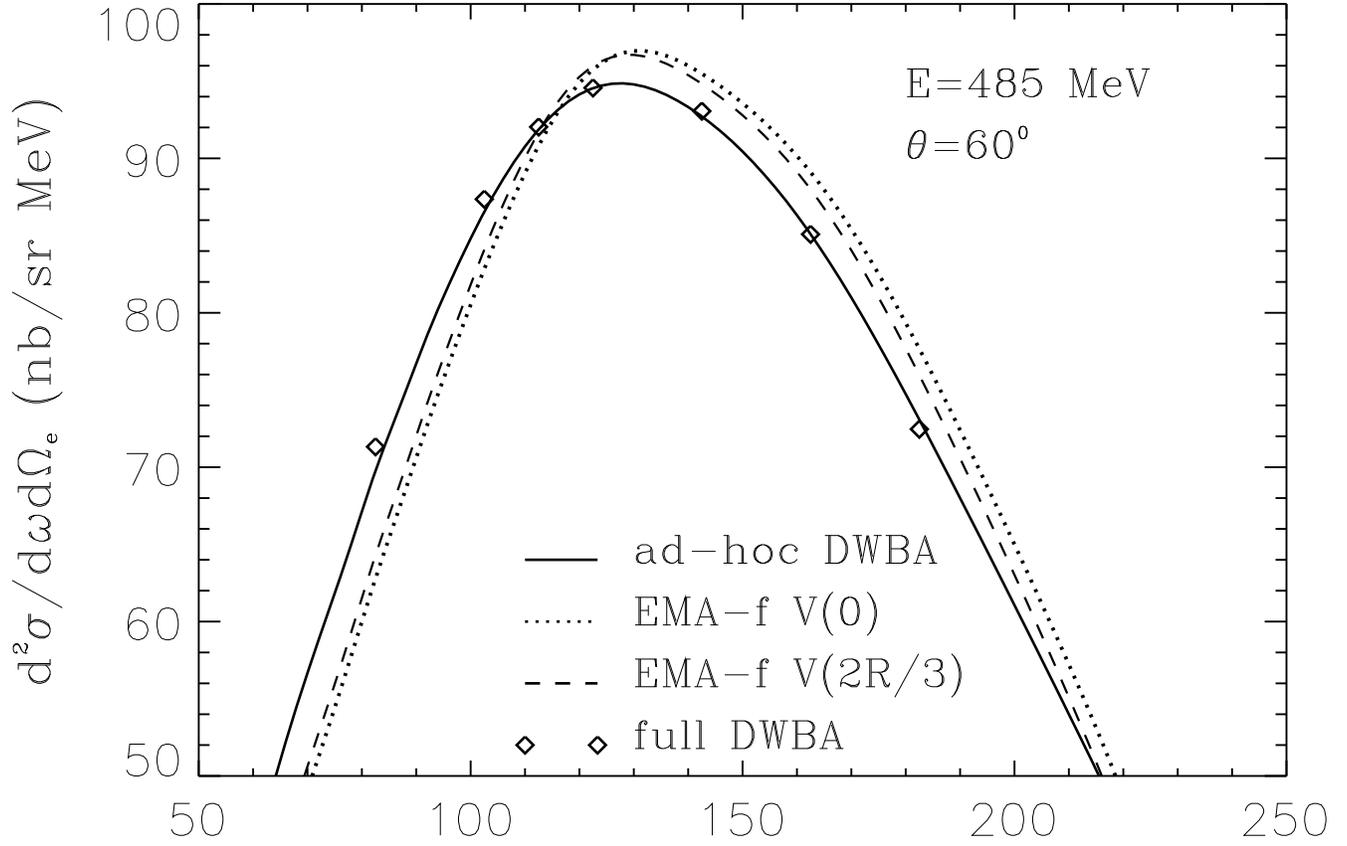} 
\caption{Combined longitudinal and transverse contributions for the 
kinematics described in Fig. \ref{ema310}.}
\label{ema485}
\end{figure*}

We have investigated the validity of the EMA-f approximation for different 
choices of $R_c$, the argument of $V_c$, by calculating the longitudinal 
and transverse contributions to the cross section using a relativistic 
nuclear model that we have successfully used to describe a great deal 
of $(e,e')$ data \cite{kim4, kim5, kim6}.  
In Fig. \ref{ema310}, we compare the cross section for the inclusive 
reaction $(e,e')$ on $^{208}$Pb for an incident electron energy of 310 MeV 
at a backward angle ($\theta=143^o$).  
We show three curves in addition to diamonds representing the full DWBA 
calculation.  As we have noted before, even at these low energies, 
our {\it ad-hoc} result agrees quite well with the full DWBA results.  
We show two EMA-f results.  
In one case we evaluated the Coulomb potential energy at the origin and 
in the other at two thirds of the nuclear radius which is in better 
agreement with the full DWBA results.  
Note that the cross section at such a large angle is dominated by the
transverse contributions.  
In Fig. \ref{ss485} we compare our {\it ad-hoc} results to EMA-f for 
the same two choices of the Coulomb potential for the longitudinal 
(upper panel) and transverse (lower panel) contributions to the cross 
section for 485 MeV electron on $^{208}$Pb at a scattering angle of $60^o$.  
Note that the longitudinal and transverse contributions are of similar 
magnitude, but that the transverse cross section is much better described 
by the EMA-f than the longitudinal cross section.  
And furthermore, using the Coulomb potential at $2/3R$ to calculate $q'$ 
is a somewhat better approximation than using the origin value.  
In Fig. \ref{ema485} we show the full cross section for this case as 
compared to the full DWBA calculation (where we cannot separate out the 
longitudinal and transverse terms) and we note that the {\it ad-hoc} 
result is very good and the EMA-f does not look so bad since over half the 
cross section comes from the transverse term which is reasonably well 
described by EMA-f.  
Of course if you are using EMA-f to make a Rosenbluth separation, 
the differential quality of the description of the longitudinal and 
transverse terms would lead to large errors.

\begin{figure*} 
\includegraphics{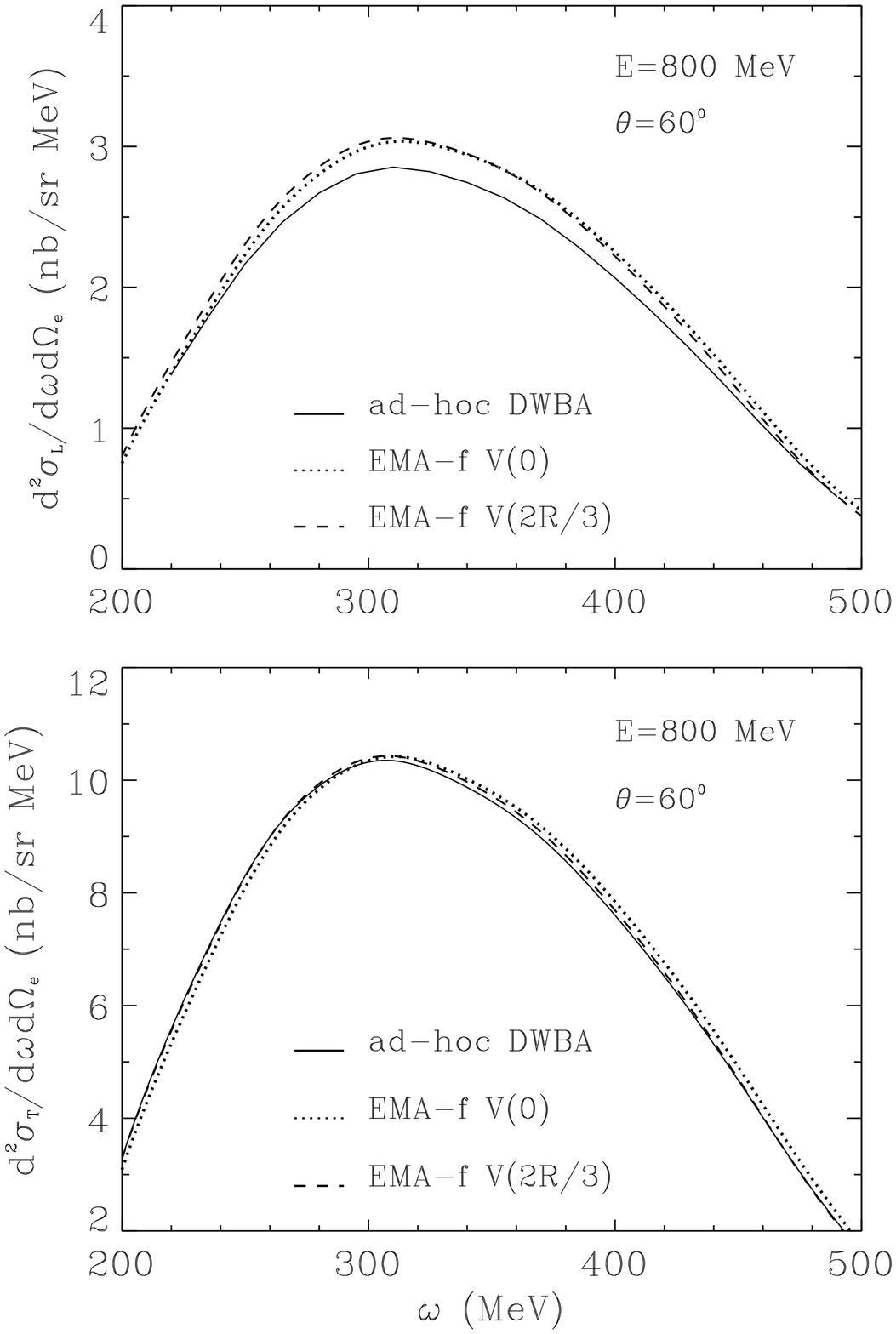} 
\caption{The quasielastic longitudinal and transverse contributions to 
the cross section for $^{208}$Pb with incident electron energy of 
$E_i$=800 MeV and electron scattering angle $\theta=60^o$ calculated 
with the {\it ad-hoc} DWBA as compared to the EMA-f with two different 
values of the Coulomb potential energy. }
\label{ss800}
\end{figure*}

\begin{figure*} 
\includegraphics{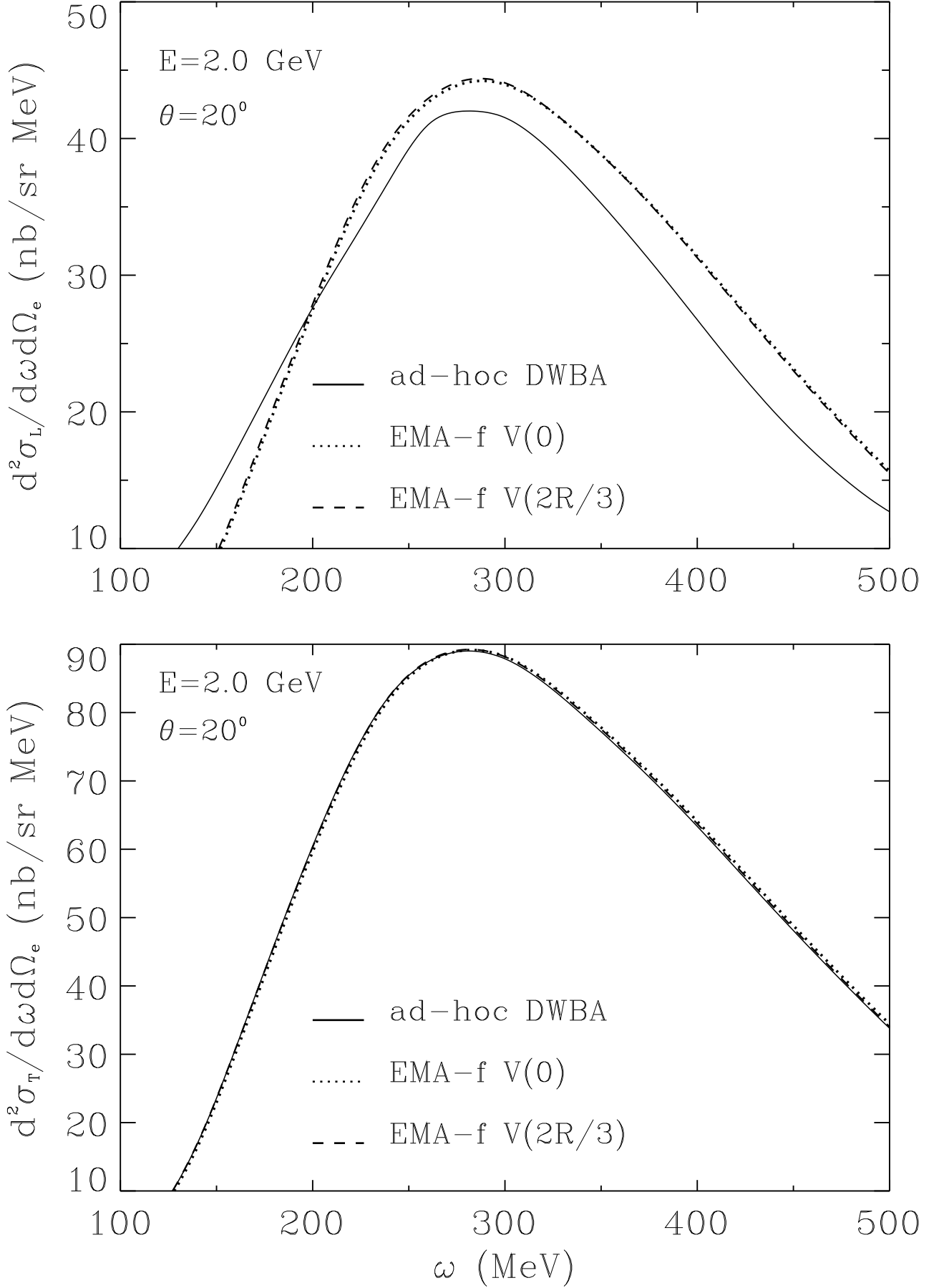} 
\caption{Same as Fig. \ref{ss800} except that $E_i$=2.0 GeV and the electron 
scattering angle $\theta=20^o$.}
\label{ss2000}\end{figure*}

In Fig. \ref{ss800}, we examine the two separated contributions as 
in Fig. \ref{ss485}, except that we have increased the electron energy 
up to 800 MeV.  
Clearly the EMA-f approximation is much better at this higher energy 
although the discrepancy in the longitudinal case is still almost 10\% 
at the quasielastic peak.  
In Fig. \ref{ss2000}, we increased the electron energy up to 2 GeV 
while reducing the scattering angle to $20^o$.  
In all cases we use $^{208}$Pb as the target.  
The EMA-f is almost exact for the transverse contribution to the cross 
section (lower panel) while the longitudinal contribution (upper panel) 
continues to have problems.  

\section{Proposed Solution and Conclusions}

\begin{figure*} 
\includegraphics{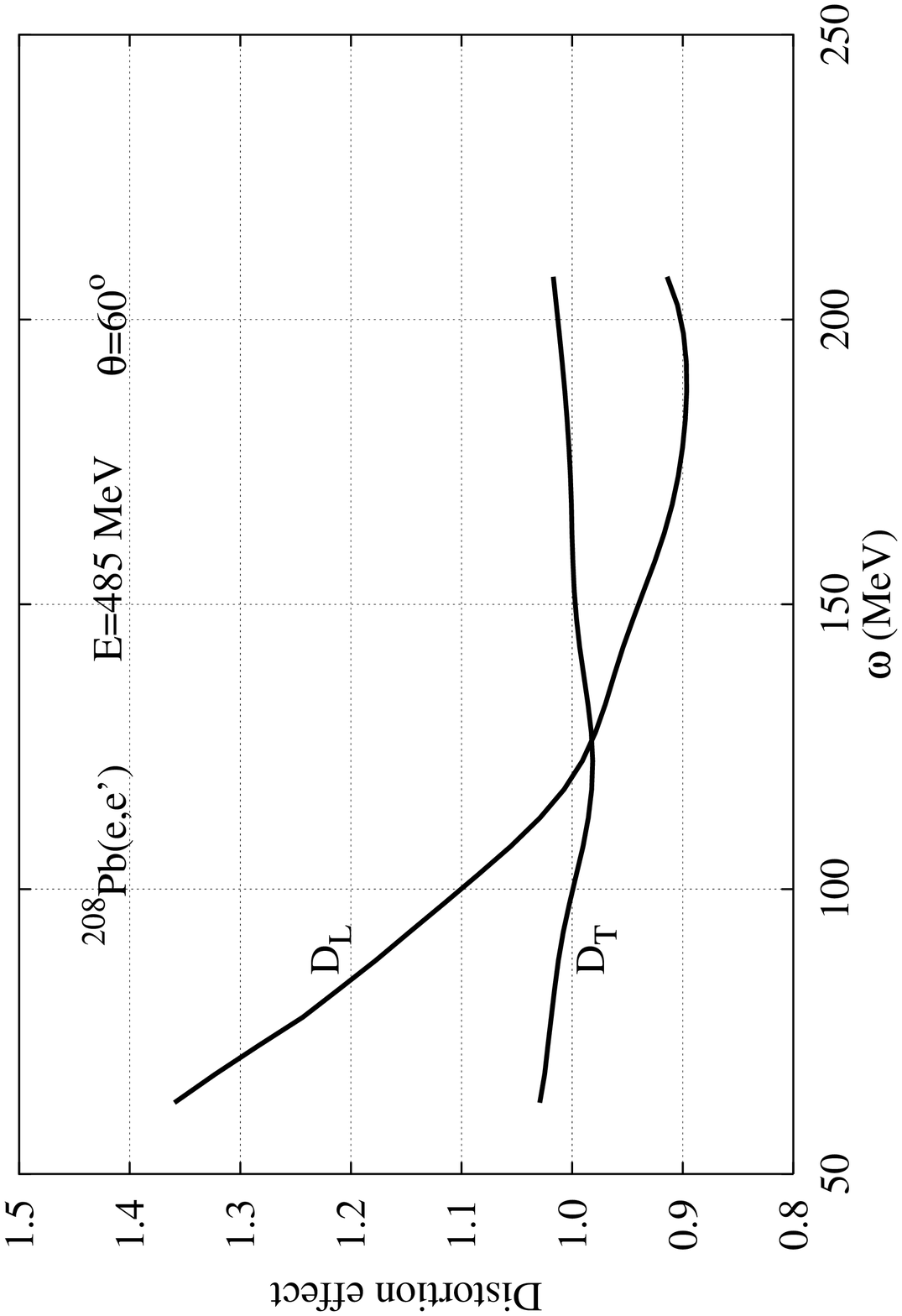} 
\caption{The ratio of the longitudinal and transverse ($D_L$ and $D_T$) 
contributions to the quasielastic cross section calculated using
the {\it ad-hoc} DWBA  method of including Coulomb effects and the EMA-f 
method of approximating Coulomb distortion for 485 MeV electrons on 
$^{208}$Pb at a scattering angle of $\theta=60^o$}.
\label{dis485}
\end{figure*}

\begin{figure*} 
\includegraphics{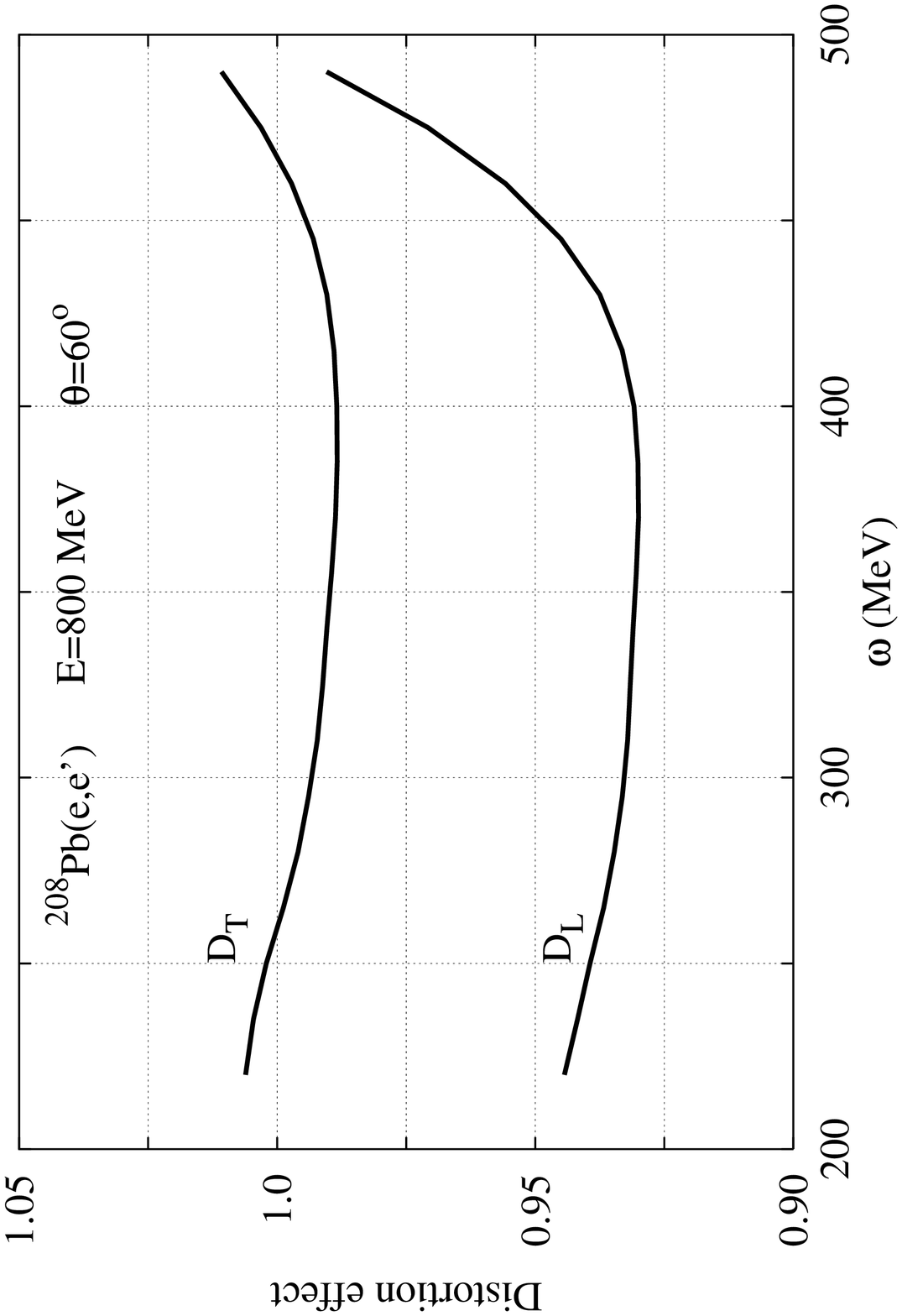} 
\caption{The ratio of the longitudinal and transverse  D$_T$ contributions to the quasielastic cross section calculated using
the {\it ad-hoc} DWBA  method of including Coulomb effects and the EMA-f method of approximating Coulomb distortion for 800 MeV electrons on $^{208}$Pb at a scattering angle of $\theta=60^o$}.
\label{dis800}
\end{figure*}

Our results show that even at rather high electron energies, the longitudinal 
contributions to the quasielastic cross section are not well described by 
the EMA approximation even with a {\it focusing} factor.  
Based on these results, we do not believe that a model-independent EMA-like 
approach can be used to extract the longitudinal structure function in 
quasielastic scattering.  
However, as the electron energy increases the EMA-f approach does get 
better and better so perhaps it can be used as the basic analysis tool, 
but with some model dependent corrections to the Coulomb distortion effects.  
In Fig. \ref{dis485}  we show the ratio of the contributions to the 
quasielastic cross section for 485 MeV electrons on $^{208}$Pb at a 
scattering angle of $60^o$ calculated using our {\it ad-hoc} model over 
the EMA-f calculation using the Coulomb potential at $\frac{2}{3}$R for the 
longitudinal and for the transverse contributions.  
As was clear in Fig. \ref{ss485}, the distortion factor $D_T$ for the 
transverse contributions differs from one by only a few percent over the 
quasielastic peak.  
However, the distortion factor $D_L$ varies considerably from one at this 
energy.   
In Fig. \ref{dis800} we repeat this calculation for incident electrons of 
800 MeV.  
At this higher energy, $D_T$ differs from one by less than 3\% across 
the quasielastic peak while $D_L$ deviates from one by up to about 7\%.   
Based on these results and our results at higher energies, we propose 
that the quasielastic scattering cross section for $(e,e')$ be written as,
\begin{eqnarray}
\left({\frac {d^2\sigma} {d\Omega_e d\omega}}\right)= 
 {\sigma_M} \left\{ 
{\frac {{Q'}^4} {{q'}^4}}  D_LS_L (q',\omega) + 
\left[ \tan^2 {\frac {\theta_e} {2}} + {\frac {{Q'}^2} {2{q'}^2}} \right]  
D_TS_T (q',\omega) \right\}. 
\label{CoulombCor}
\end{eqnarray}
where the distortion factors D$_L$ and D$_T$ are given by
\begin{eqnarray}
D_L&=&\left(\frac{Q}{Q'}\right)^4 \left(\frac{q'}{q}\right)^4 
\frac{S'_L(q',w)}{S_L(q',\omega)} \\ 
D_T&=&\left(\frac{p'_i(0)}{p_i}\right)^2 \frac{S'_T(q',\omega)}
{S_T(q',\omega)}
\end{eqnarray}
Note that in the factor before the transverse structure function in Eq. (\ref{CoulombCor}), we are using the factor $Q^2/2q^2$ rather than primed values so as to more easily define the distortion factor $D_T$.
Clearly by using distortion factors $D_L$ and $D_T$ calculated from any 
reasonable nuclear model, Eq. (\ref{CoulombCor}) can be used in a 
Rosenbluth mode to extract the longitudinal and transverse functions as 
a function of q$'$ and $\omega$ from experimental data.  

In conclusion, we have shown that the effective momentum approximation 
(using the Coulomb potential at $\frac{2}{3}$R) with an overall 
{\it focusing} factor of $\left(p'_i(0)/p_i\right)^2$ is a very good 
approximation of the Coulomb distortion effects for the transverse 
contributions to the quasielastic cross section.  
However, for electron energies less that about 600 MeV it is not a 
good approximation of the longitudinal contributions.  
At higher electron energies this approximation does get better for 
the longitudinal contribution, but for accurate extraction of the 
longitudinal structure function it is necessary to use distortion 
factors calculated from a nuclear model.  
The procedure we have proposed minimizes the model dependence by only 
using the model to evaluate the Coulomb distortion effects not included 
in EMA-f.   
Based on the cases we have examined, we believe the errors in the 
extracted structure functions arising from Coulomb distortion effects 
using our proposed procedure should be less than 5\% for incident 
electron energies above 600 MeV.

\section*{Acknowledgments}
This work was supported the US Department
of Energy under Contract DE-FG02-93ER40756 with Ohio University.

\end{document}